\documentclass[pdflatex,sn-mathphys-num]{sn-jnl}

\usepackage{graphicx}%
\usepackage{multirow}%
\usepackage{amsmath,amssymb,amsfonts}%
\usepackage{amsthm}%
\usepackage{mathrsfs}%
\usepackage[title]{appendix}%
\usepackage[dvipsnames,svgnames,table]{xcolor}
\usepackage{textcomp}%
\usepackage{bold-extra}
\usepackage{manyfoot}%
\usepackage[linesnumbered,ruled,vlined]{algorithm2e}
\usepackage{booktabs}%

\usepackage{listings}%
\usepackage{xparse}
\usepackage{pythonhighlight}

\theoremstyle{thmstyleone}%
\theoremstyle{thmstyletwo}%

\theoremstyle{thmstylethree}%

\definecolor{lightgray}{gray}{0.95}
\let\oldtabular\tabular
\let\endoldtabular\endtabular
\renewenvironment{tabular}{\rowcolors{2}{white}{lightgray}\oldtabular}{\endoldtabular}

\usepackage{xcolor}

\usepackage{lineno}

\begin{document}

\title[Article Title]{Fulqrum: a generalized framework for quantum subspace diagonalization}


\author*[1]{\fnm{Paul D.} \sur{Nation}}\email{paul.nation@ibm.com}
\equalcont{These authors contributed equally to this work.}

\author[2]{\fnm{Abdullah Ash} \sur{Saki}}
\equalcont{These authors contributed equally to this work.}

\author[1]{\fnm{Hwajung} \sur{Kang}}

\affil[1]{\orgdiv{IBM Quantum}, \orgname{T. J. Watson Research Center}, \orgaddress{\city{Yorktown Heights}, \postcode{10598}, \state{NY}, \country{USA}}}

\affil[2]{\orgdiv{IBM Quantum}, \orgname{IBM Research Cambridge}, \orgaddress{\city{Cambridge}, \postcode{02142}, \state{MA}, \country{USA}}}

\abstract{We present an open-source software framework for computing the solution to Hamiltonian eigenproblems in a subspace defined by bit-strings sampled from a quantum computer.  Hamiltonians are represented using an extended alphabet that includes projection and ladder operators, yielding a unified solution method for qubit and fermionic systems.  Operators are grouped and sorted so that only non-zero terms need be evaluated and a minimal number of subspace lookup operations performed. Bit-strings are expressed using bit-sets to reduce memory consumption and allow for evaluating operators with no fundamental limit on the number of qubits. Subspaces defined over bit-sets are stored in a hash map format that allows for efficient indexing and lookup operations. Specializations for problems, such as those from quantum chemistry, can further reduce computational costs. Our method can be used to directly construct sparse matrix Hamiltonian representations or obtain matrix-free solutions, allowing users to utilize their eigensolver of choice.  We show the benefits of our framework by computing the ground-state solution to examples from condensed matter physics and quantum chemistry with less memory and runtime compared to existing techniques, in some cases by two orders of magnitude or more.  This work provides a flexible interface for performant quantum-classical eigensolutions for candidate quantum advantage applications.}


\maketitle
Eigenvalue problems represent one of the mainstay calculations for near-term \cite{bharti:2022, cerezo:2021} and fault-tolerant \cite{kitaev:1995, abrams:1999, poulin:2009, dong:2022, alexeev:2025} quantum applications, including domains such as quantum chemistry \cite{mcardle:2020, motta:2023} and quantum optimization \cite{abbas:2024}.  Recent demonstrations have shown that such techniques offer promise towards achieving quantum advantage using pre-fault tolerant quantum hardware \cite{kanno:2023,nakagawa:2023,robledo:2024,sugisaki:2024,pellow:2025,yu:2025, smith:2025, barroca:2025, barison:2025, danilov:2025, shirakawa:2025, wang:2025, piccinelli:2026, li:2026, merz:2026, das:2026}. Here, eigensolutions are formulated within a subspace defined by bit-strings sampled from a quantum computer and leverage classical computing resources to perform the eigenvalue computation itself. 

These methods aim to solve Hermitian eigenproblems of the form $\hat{H}|\psi_{j}\rangle = \lambda_{j} |\psi_{j}\rangle$, or the generalized version $\hat{H}|\psi_{j}\rangle = \lambda_{j} M|\psi_{j}\rangle$, where $\hat{H}$ is the Hamiltonian of a qubit or fermionic system, $\lambda_{j}$ and $|\psi_{j}\rangle$ are the target eigenvalue and eigenvector, respectively, and $M$ is a Hermitian ``overlap" matrix \cite{epperly:2022}.  The state $|\psi_{j}\rangle$ is defined in a subspace spanned by the counts returned from sampling one or more quantum circuits on a quantum processor.  Implicit in these methods is that $|\psi_{j}\rangle$ has compact support over the Hilbert space, allowing for accurate sampling with only a polynomial number of circuit repetitions.  Any possible quantum advantage requires sampling bit-strings pertinent to the target eigenstate faster on a quantum computer than counterpart classical algorithms.  With a Hamiltonian represented in the computational basis, such problems can readily be cast as sparse matrix eigenproblems that are amenable to a variety of classical solution methods such as \textsc{ARPACK} \cite{arpack}, \textsc{PRIMME} \cite{primme}, and \textsc{SLEPc} \cite{slepc}.  At their core, these eigensolution methods require repeatedly solving sparse matrix-vector products (SpMV), and the conversion of Hamiltonian operators and subspaces of bit-strings into data structures compatible with efficient SpMV operations is a key step in these workflows \cite{saad:2011}.  Because only the output vector from SpMV is utilized, matrix-free methods that require no explicit matrix storage can also be used provided computing individual matrix-elements is efficient.

Currently, hybrid quantum-classical Sample-based Quantum Diagonalization (SQD) approaches are bottle-necked by the runtime associated with the classical diagonalization step.  For example, the original SQD paper, Ref.~\cite{robledo:2024}, took $\sim 45$ minutes of quantum processor (QPU) time to sample the circuit for $\rm Fe_{4}\rm S_{4}$, and $96$ CPU node-hours ($4608$ core-hours) of classical runtime per diagonalization on the Fugaku supercomputer.  Typically multiple diagonalizations are required as the subspace is processed in batches or iteratively improved, further skewing the quantum / classical runtime imbalance.  More recent work utilizing GPU resources, Ref.~\cite{das:2026}, shows similar behavior with net QPU run times of $9$ hours and classical diagonalization resources of $327$ GPU node-hours.  As such, any improvements into the runtime (and memory consumption) of SpMV operations promises to greatly accelerate these applications and allow for extending their computational reach to larger problem instances.

In general, SQD software frameworks are tied to a domain-specific application area and are not generally applicable to other problem instances.  For example, quantum chemistry has a well-versed set of tools for solving the eigenstates of fermionic systems \cite{pyscf, dice, psi4, sbd}, but that are incapable of addressing qubit Hamiltonians.  Conversely, tools for solving qubit based systems also exist \cite{sqd} and can solve fermionic systems as well, provided one performs a fermionic to qubit Hamiltonian transformation, e.g. a Jordan-Wigner (JW) transformation \cite{jordan:1928}, but otherwise use independent solution methods and input / output formats; one cannot substitute eigensolvers without substantially modifying the overall workflow. In addition, many of these SQD solvers are limited to the length of bit-strings that they support due to restrictions representing bit-strings with integer data types.  For example, \texttt{qiskit-addon-sqd} handles at most $63$ qubits for spin systems, while \textsc{PySCF} \cite{pyscf-repo} and \textsc{Dice} \cite{dice} support $<=63$ orbitals, or bit-strings of maximum length $126$.  In contrast, newer tools like \textsc{SBD} \cite{sbd} (a modern \textsc{Dice} derivative) support bit-strings of arbitrary length.  With quantum computing hardware already fielding qubit counts of $\mathcal{O}(100)$ or more, having no strict limit on the size of supported bit-strings is beneficial for solving problems at the frontier of quantum advantage and beyond.  Note that these limitations are not fundamental, but rather arise due to the data structures used to encode the problem instance.

\begin{figure*}[t]
\centering
\includegraphics[width=\textwidth]{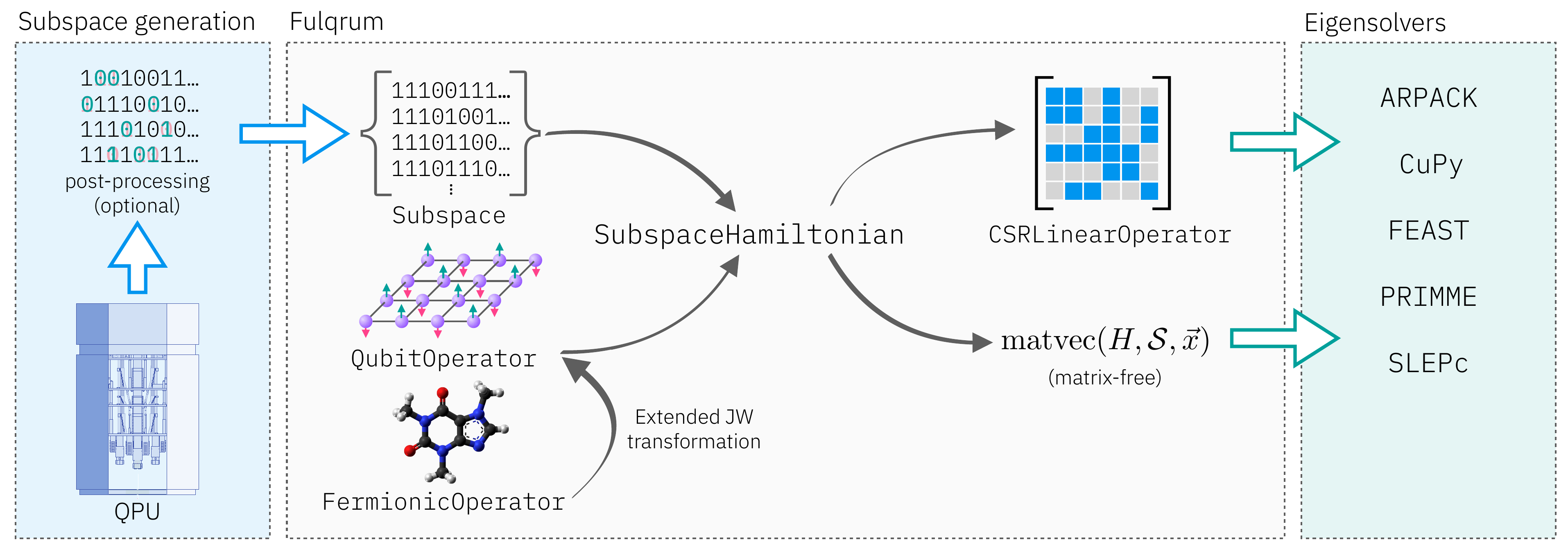}
  \caption{Exemplar \textsc{Fulqrum} workflow.  Bit-strings sampled from a quantum processing unit (QPU), with optional post-processing such as configuration recovery \cite{robledo:2024}, are used to construct a \texttt{Subspace} instance that, together with a corresponding qubit (\texttt{QubitOperator}) or fermionic (\texttt{FermionicOperator}) Hamiltonian representation, are used to build a \texttt{SubspaceHamiltian} model.  Fermionic systems are cast to equivalent qubit representations via an extended-alphabet JW transformation for a unified solution method.  From a \texttt{SubspaceHamiltonian}, users can generate CSR-matrix representations (\texttt{CSRLinearOperator}), or utilize matrix-free solution methods via the abstract \texttt{LinearOperator} class (matvec) in SciPy, passing these on to a wide range of classical eigensolvers \cite{arpack, cupy, polizzi:2009, primme, slepc}.}\label{fig:workflow}
\end{figure*}

In this work we present a generalized framework, called \textsc{Fulqrum} \cite{fulqrum}, for performing \textit{both} qubit and fermionic SQD.  This is accomplished by efficiently casting fermionic problems to qubit Hamiltonians consisting of an operator alphabet that includes an extended set of projection and ladder operators. In addition, caching optimizations for quantum chemistry derived problem instances yield noticable performance advantages over current techniques.  Utilizing bit-sets for computational basis states, as opposed to integers, our method has no fundamental limit on the number of allowed qubits.  \textsc{Fulqrum} does not perform eigensolving itself, instead generating canonical SpMV representations that are compatible with a wide range of existing numerical methods, giving users a choice of classical eigensolver and related tools such as preconditioners and matrix-reordering techniques leveraging both CPUs and GPUs. An overview of this workflow is presented in Fig.~(\ref{fig:workflow}).  Although generalized solution techniques often come at the expense of performance, in terms of runtime, memory consumption, or both, here we will demonstrate that it is possible to outperform current implementations in terms of both metrics whilst maintaining flexibility.

\section{Results}\label{sec:results}

Here we utilize two computing platforms: ``Computer A" is a 12-core, 24-thread AMD 7900 processor with 128~GiB of memory running Ubuntu 26.04, while ``Computer B" is a dual-socket Intel Xeon Platinum 8474C (48-core, 96 threads) workstation with 2~TiB of memory running Ubuntu 22.04.  In this section we highlight the flexibility of our framework and validate its performance by presenting results for solving the ground state eigensolution for both qubit and fermionic Hamiltonians with CPU resources.  Using \textsc{Fulqrum} with GPU accelerators will be discussed in a companion paper.  Note that the results presented here focus on relative comparisons with other SQD eigensolution methods in terms of runtime and memory consumption rather than accuracy against a known solution.  The ability to obtain accurate solutions is dictated by the sampled bit-strings that comprise the subspace and that are gathered independently from the diagonalization procedure itself; the subspace is an input to the eigensolving process and all solvers should return numerically similar answers.

In Fig.~(\ref{fig:code}) we show a simple quantum chemistry workflow using the \textsc{Fulqrum} API.  In this example, a molecule defined by an \textsc{FCIDump} \cite{fcidump} file along with a collection of bit-strings is used to construct a subspace qubit Hamiltonian representation that can be used for SpMV operations and passed onto supporting eigensolvers.  The steps of the code mirrors the workflow presented in Fig.~(\ref{fig:workflow}).  All examples in this section follow this same general program structure.

\begin{figure}[h]
\centering
\begin{python}
import primme
import fulqrum as fq

# Load operator from FCIDump file
fop = fq.FermionicOperator.from_fcidump("fcidump.txt")
# Perform extended JW transformation to qubit operator
op = fop.extended_jw_transformation()
# Build TPS subspace from given bit-strings
S = fq.Subspace([alpha_strs, beta_strs])

# SubspaceHamiltonian object from operator and subspace
Hsub = fq.SubspaceHamiltonian(op, S)

# Perform matrix-free eigensolving using PRIMME
eigvals, eigvecs = primme.eigsh(Hsub, k=1, which="SA")
\end{python}
\caption{Sample quantum chemistry workflow in \textsc{Fulqrum} taking a molecule in a FCIDump file along with a collection of bit-strings sampled on a quantum computer, or classical generation technique, and returning the resulting ground state energy using the \textsc{PRIMME} eigensolver.}
\label{fig:code}
\end{figure}

\subsection{Condensed matter}
To begin, we consider the following antiferromagnetic XXZ spin-1/2 chain Heisenberg model, 
\begin{align*} 
\label{eq:heisen}
H = \sum^{L-1}_{i=1} J(X_iX_{i+1} + Y_iY_{i+1}) + Z_iZ_{i+1},
\end{align*}
where $L$ is the number of sites and $J$ is the coupling strength for the off-diagonal term, $X_iX_{i+1} + Y_iY_{i+1}$, and we set $J=0.3$.  To estimate the ground state energy for this Hamiltonian, we utilize Sample-based Krylov Quantum Diagonalization (SKQD) \cite{yu:2025}.  SKQD is a SQD algorithm for near-term quantum computers where the subspace is comprised of quantum Krylov states built from a set of time-evolved states generated from a reference state using the target Hamiltonian.  Upon sampling the Krylov states, diagonalization of the Hamiltonian in the generated subspace is performed classically.  We construct fifteen quantum circuits corresponding to different time steps incremented by $\Delta t = 0.1$ (trotter steps), starting from zero, for each $L \in \{30, 36, 40, 46, 50, 56, 60\}$, and evolve them from a reference Neel state, $|01\rangle^{\otimes L/2}$ on the 156-qubit IBM Marrakesh Heron device.  The minimum chain length is set such that meaningful timing information can be gathered, while the upper bound comes from the 63-qubit limit of the \texttt{qiskit-addon-sqd} \cite{sqd} qubit Hamiltonian eigensolver that we compare against. To account for errors, we expand the subspace by post-processing sampled bit-strings that are a Hamming distance one away from the Neel state, exploiting conservation of total magnetization in this model, keeping all bit-strings where a single bit-flip gives the correct total magnetization. All other bit-strings are discarded from the subspace.  Using these corrected bit-strings, we compare the performance of \textsc{Fulqrum} and \texttt{qiskit-addon-sqd} for projection of the Hamiltonian into the subspace, as well as classical eigensolving.  Here we use \textsc{Fqulrum}'s ``fast" CSR matrix building mode, see Sec.~(\ref{sec:evaluation}). For the diagonalization procedure, the \textsc{SciPy} \cite{scipy} eigensolver \textsc{ARPACK} \cite{arpack} is used for both methods as this choice is hard-coded in the \texttt{qiskit-addon-sqd} qubit solver.  In contrast, \textsc{Fulqrum} allows the user to select their eigensolver of choice, potentially enabling more performant eigensolving as the method can be targeted to the Hamiltonian under consideration.  Here a uniform vector is used as the initial vector passed to the eigensolver for both frameworks.

\begin{figure}[t]
    \centering
    \includegraphics[width=10cm]{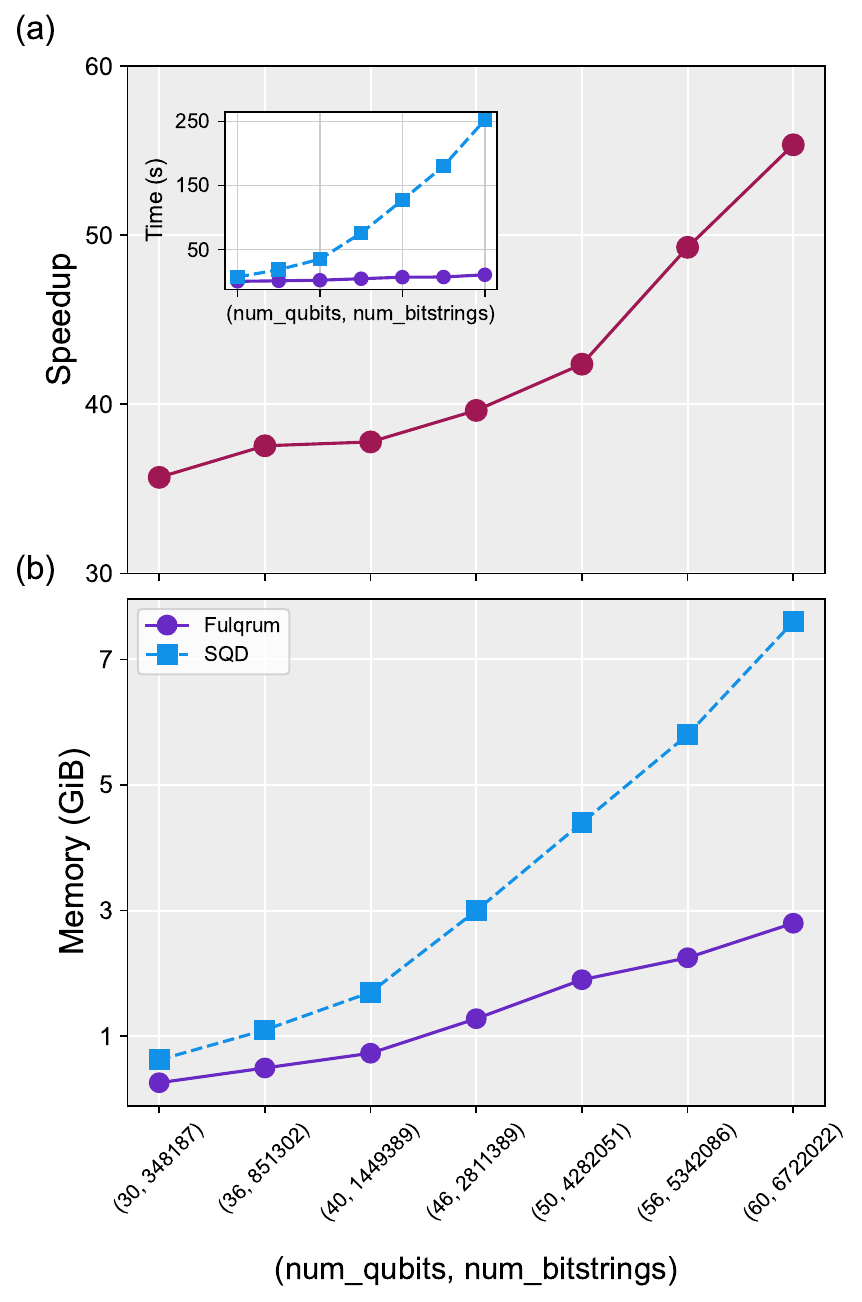}
    \caption{(a) Ratio of \texttt{qiskit-addon-sqd} (\textsc{SQD}) to \textsc{Fulqrum} runtimes for projecting a 1D Heisenberg spin chain Hamiltonian over subspaces defined by post-processed bit-strings sampled from the IBM Marrakesh device for differing numbers of qubits running on Computer A.  Inset shows time to solution for the full ground state eigenproblem of the same Hamiltonians and subspaces using \textsc{qiskit-addon-sqd} (squares-dashed) and \textsc{Fulqrum} (circles-solid). The lowest of five runs is used for timing. (b) Peak memory consumption for full eigensolutions to problems presented in (a).}
    \label{fig:spin-results}
\end{figure}

The SKQD benchmark is executed on Computer A and evaluated in terms of runtime and peak-memory usage.  As the core differentiating routine is projection of the Hamiltonian into the target subspace, we focus on comparing the time taken in this step. However, the peak memory usage is measured over the full eigensolving workflow so as to faithfully capture the resources required to solve the complete problem.  Results are displayed in Fig.~(\ref{fig:spin-results}) where we see that even at the smallest number of qubits and bit-strings examined by \textsc{Fulqrum} is over an order of magnitude faster than \texttt{qiskit-addon-sqd}, with the difference in projection timings diverging as the number of qubits and bit-strings grows; \textsc{Fulqrum} scales markedly better than \texttt{qiskit-addon-sqd} as problem sizes extend into the regime where quantum applications may show advantage over their classical counterparts.  A similar trend is seen for the difference in timing of the full eigen workflow, inset of Fig.~(\ref{fig:spin-results}a), where, although both solvers make use of \textsc{ARPACK}, unlike \texttt{qiskit-addon-sqd} that uses the default serial SpMV in \textsc{SciPy}, \textsc{Fulqrum} wraps its SpMV implementation in a \texttt{CSRLinearOperator} [see Sec.~(\ref{sec:evaluation})] allowing for processing matrix rows in parallel using \textsc{OpenMP}.  Finally, the difference between the peak memory consumption of the two implementations also widens with subspace dimension and number of qubits, allowing \textsc{Fulqrum} to solve larger problem instances given fixed memory resources.  Note that, while we could also use a matrix-free solution, the memory size of the resultant matrix is much smaller than that needed to store the subspace information, and thus is it not practical to do so.  This is typical for problems, such as the one presented here, that are defined on low-degree graphs.

\subsection{Quantum chemistry}
For fermionic Hamiltonians we select four molecules.  First is the $\mathrm{N}_{2}$ molecule in the cc-pvdz basis \cite{dunning:1989}, second a methane ($\mathrm{CH}_{4}$) dimer in the aug-cc-pVQZ basis \cite{kaliakin:2025}, next is $\mathrm{Fe}_{4}\mathrm{S}_{4}$ expressed in the TZP-DKH basis \cite{robledo:2024}, and finally Benzene ($\rm C_{6}\rm H_{6}$) in the 6-31G basis. For the $\mathrm{N}_{2}$ molecule, we generate bit-strings by running the corresponding chemistry-specific Local Unitary Cluster Jastrow (LUCJ)~\cite{lucj} quantum circuit on the IBM Boston 156-qubit Heron quantum processor, sampling the circuit one million times. Out of these samples, $3,722$ are bit-strings generated with the correct electron number in each spin sector. To correct the remaining bit-strings we run one round of configuration recovery to generate bit-strings with physically valid numbers of electrons \cite{robledo:2024}. The full corrected set of bit-string is split into alpha (spin up) and beta (spin down) halves and curated together with duplicates removed. Finally, the first $6,000$ lexicographically sorted half bit-strings from this collection are selected to span the final subspace of $36$ million full bit-strings by taking the tensor-product; we generate a Tensor-Product Subspace (TPS).  The sampled quantum subspace for the $\mathrm{CH}_{4}$ dimer is a TPS comprised of configuration recovered bit-strings taken from Ref.~\cite{kaliakin:2025}.  The computational benefits of using a TPS is discussed in Sec.~(\ref{sec:subspace}).  

Subspaces for $\rm Fe_{4} \rm S_{4}$ and Benzene were generated using a recursive perturbative expansion around the Hartree-Fock (HF) state.  A discussion of this ``RAMPS" technique is presented in Sec.~(\ref{sec:ramps}). In the case of $\rm Fe_{4} \rm S_{4}$, RAMPS is run through a maximum of $3$ recursions with a tolerance of $10^{-14}$ generating $\sim 42$ million bit-strings.  From this, the first unique $10,000$ half bit-strings are used for generating a TPS up to a dimension of $10^{8}$. The TPS used for Benzene is constructed in an iterative manner using two-rounds of the RAMPS procedure, each performing a single-recursion at a tolerance of $10^{-8}$, with the bit-strings and ground state energy from the first round used as the initial subspace and energy, respectively, for the second call.  This yields $13,376$ half bit-strings for a TPS.  The difference in subspace construction methods presented here arises from explorations into how best to utilize RAMPS for subspace construction and can likely be improved upon when using initial subspaces beyond HF.   

Detailed molecular information, subspace dimensions, and computed electronic energies for the sampled subspaces are given in Tbl.~(\ref{tab:chem-params}).  Note that the product of the subspace dimension and the number of groups equals the number of matrix-element evaluations that needs to be performed.  However, when confined to a subspace, it does not equal the number of elements in the resultant Hamiltonian matrix.

\begin{table}[t]
\begin{tabular}{lcccc}
Parameter name & $\mathrm{N}_{2}$ & $\mathrm{CH}_{4}$ dimer & $\mathrm{Fe}_{4}\mathrm{S}_{4}$ & Benzene ($\rm C_{6} \rm H_{6}$)  \\
\hline
Num. (spatial) orbitals               & 26         & 24          &  36                          &  66\\
Num. electrons ($\alpha$/$\beta$)    & 10 (5/5)    & 16 (8/8)      & 54 (27/27) & 42 (21/21)         \\
Subspace dimension(s)               & $36,000,000$  & $65,983,129$ & $10^{4}\rightarrow 10^{8}$  & $10^{8}$  \\
Num. groups (\textsc{Fulqrum})                 & $16,073$          & $30,105$ & $515,917$ & $3,486,310$       \\
Est. electronic energy(s) (Ha.)   & $-32.89242$   & $-44.05140$ & $-326.59523 \rightarrow -326.81104 $   & $-230.39872$ \\
\end{tabular}
\caption{Parameters defining the $\mathrm{N}_{2}$, $\mathrm{CH}_{4}$-dimer, $\mathrm{Fe}_{4}\mathrm{S}_{4}$, and Benzene quantum chemistry problems considered in this work.  The width of operators in the qubit Hamiltonian generated by the extended JW transformation is twice the number of orbitals.  The estimated electronic energies are those obtained from all the eigensolvers for each of the subspaces.}
\label{tab:chem-params}
\end{table}

We compute the ground state energies for all molecules using \textsc{Fulqrum}, the popular chemistry package \textsc{Dice} \cite{dice2, dice}, and \textsc{SBD} \cite{sbd} on Computer B, comparing time-to-solution and peak memory consumption \cite{memprof}.  As \textsc{Fulqrum} does not perform eigensolving itself, we are free to select the Hermitian eigensolver \textsc{PRIMME}~\cite{primme} utilizing the \texttt{PRIMME\_LOBPCG\_OrthoBasis} method; \textsc{PRIMME}'s implementation of Locally Optimal Block Preconditioned Conjugate Gradient (LOBPCG) algorithm \cite{knyazev:2001}.  For \textsc{Fulqrum} we select an initial trial vector that has a single spike ($= 1.0$) at the index corresponding to the diagonal element within the subspace with the lowest energy (usually Hartree-Fock) and use a shifted Jacobi (diagonal) preconditioner, where the energy shift is an approximation to the lowest eigenenegy and is computed dynamically by \textsc{PRIMME}.  \textsc{Dice} is installed using the RIKEN branch from Ref.~\cite{dice-fork} and invoked using the \texttt{solve\_hci()} method from the \texttt{qiskit-addon-dice-solver}~\cite{addon-dice-solver}. \textsc{Dice} is run in \emph{selected configuration interaction} (SCI) mode that prevents the inclusion of configurations (bit-strings) outside the initial subspace by setting the maximum number of heat-bath configuration (HCI) iterations (\texttt{max\_iter}) to $1$ and the \texttt{select\_cutoff} to the largest signed 32-bit integer value ($2^{31}-1$). \textsc{SBD} is configured with \texttt{bit\_length=64}, and \texttt{num\_blocks=10}, with all other values unchanged. Each method was configured to achieve the same ground state energy given in Tbl.~(\ref{tab:chem-params}); a relative error of $\mathcal{O}(10^{-7})$. We set number of \texttt{mpirun} processes to $50$ in \textsc{Dice} runs. To ensure fairness, we set \texttt{OMP\_NUM\_THREADS=50} for \textsc{Fulqrum} and \textsc{SBD} runs for all modes. The maximum allotted memory was set to $1.5$~Tb so as not to crash the node.

\begin{figure*}[!h]
    \centering
    \includegraphics[width=.7\paperwidth]{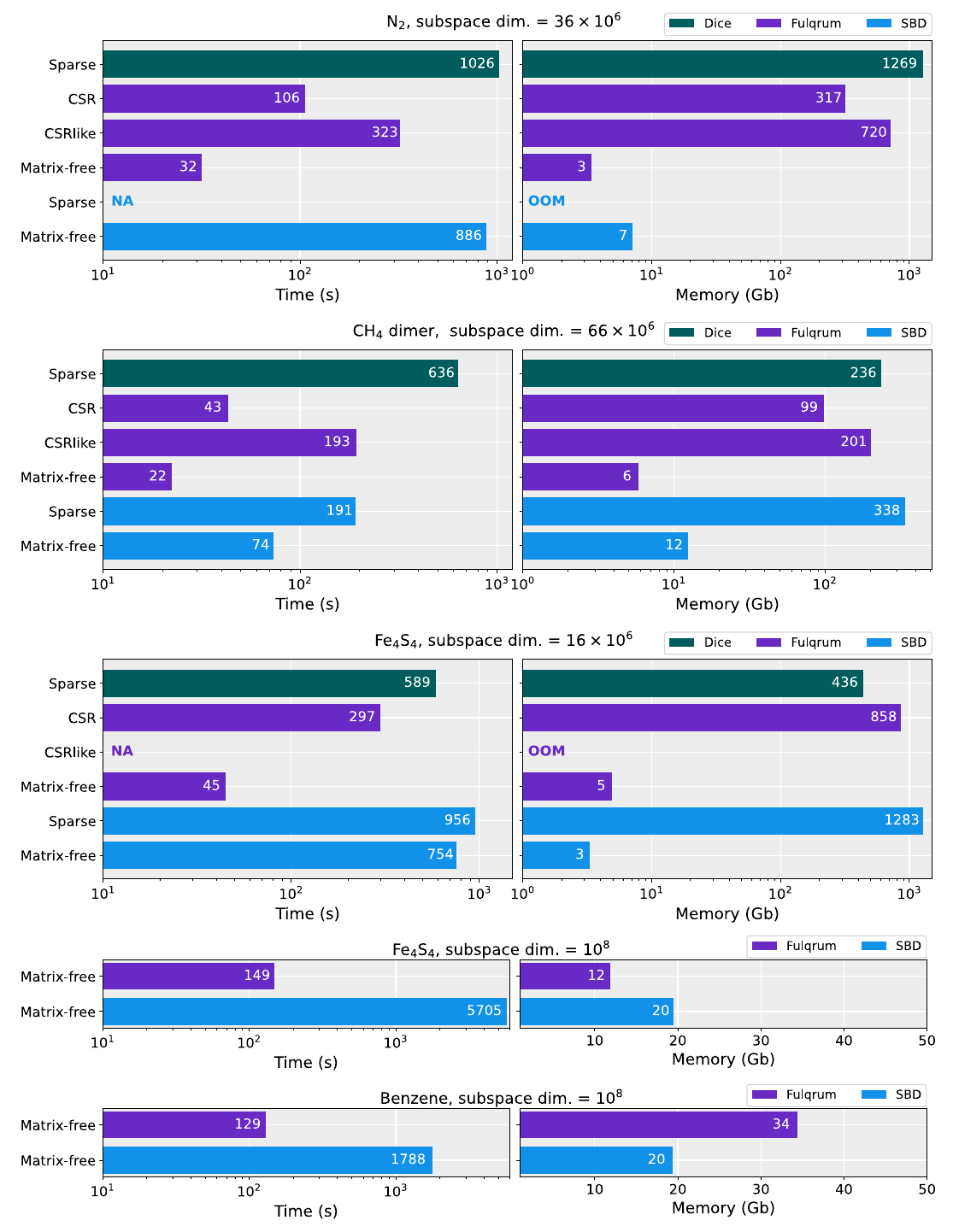}
    \caption{Total runtime and peak memory consumption for the ground state solution for the molecules in Tbl.~(\ref{tab:chem-params}) using \textsc{Dice} (green), \textsc{Fulqrum} (purple), and \textsc{SBD} (blue) running on Computer B.  \textsc{Dice} is run in ``sparse" mode that constructs a sparse structural representation of the Hamiltonian in the subspace.  \textsc{Fulqrum} in CSR, ``fast" CSR, and matrix-free modes, as described in Sec.~(\ref{sec:evaluation}).  \textsc{SBD} includes a ``sparse" mode similar to that of \textsc{Dice} as well as matrix-free evaluation.  Here, ``NA'' indicates a mode that did not complete due to an Out of Memory (OOM) error.  Times shown are the mean of three runs. Labels show the elapsed time and memory rounded to the nearest second and gigabyte, respectively, for brevity.}
    \label{fig:chemistry-results}
\end{figure*}

The results of this comparison are presented in Fig.~(\ref{fig:chemistry-results}) where it is clear that \textsc{Fulqrum} outperforms both \textsc{Dice} and \textsc{SBD} in terms of runtime when utilizing either matrix-free or CSR SpMV modes, if the latter fits into the available memory. \textsc{Fulqrum}'s matrix-free mode is particularly performant, being $3 \rm x$ to $38 \rm x$ faster than the next best competitor, depending on the molecule and subspace dimension. \textsc{SBD} tends to have faster a runtime than \textsc{Dice}, except in the case of $\rm Fe_{4} \rm S_{4}$ where the \textsc{SBD} solution times become relatively longer than the other solvers.  This is exasperated as the subspace size grows and it is seen that the matrix-free \textsc{SBD} runtime goes from $\sim 16\rm x$ greater than \textsc{Fulqrum} at $16 \times 10^{6}$ bit-strings to $\sim 38\rm x$ at a subspace dimension of $10^{8}$.

One, perhaps counterintuitive, observation is that the matrix-free modes of both \textsc{Fulqrum} and \textsc{SBD} always outperform those modes that explicitly construct a sparse representation.  While sparse representations store matrix information, matrix-free mode requires repeatedly evaluating matrix-elements for each SpMV and nominally incur a runtime overhead. This is not what we observe in Fig.~(\ref{fig:chemistry-results}).  Rather, what the results point to is memory access being a bottleneck when constructing large sparse matrices, leaving the CPU under utilized.  Matrix-free computations reduce this IO constraint and, in the case of \textsc{Fulqrum}, together with efficient caching mechanisms discussed in Sec.~(\ref{sec:fermi}), allow for maximizing SpMV performance.  This is also why \textsc{Fulqrum}'s ``fast'' CSR mode [see Sec.~(\ref{sec:evaluation})] is slower than direct CSR building for quantum chemistry problems despite working well for condensed matter and other qubit Hamiltonians; in general molecular systems contain many more terms than qubit Hamiltonians and thus are more sensitive to memory access patterns in SpMV.

Peak memory usage amounts shown in Fig.~(\ref{fig:chemistry-results}) follow the expected pattern with matrix-free evaluations requiring much less memory than explicit matrix storage.  When using a sparse representation, \textsc{Fulqrum}'s CSR mode tpyically uses the least amount of memory followed by \textsc{Dice}.  However, this is not the case for $\rm Fe_{4} \rm S_{4}$ which has many more nonzero elements than the other molecules over which \textsc{Dice} was run, as expected from the number of groups shown in Tbl.~(\ref{tab:chem-params}).  In this case, the difference in memory is due to the fact that while \textsc{Fulqrum} builds a canonical CSR matrix for SpMV, \textsc{Dice} uses a custom representation that stores only the sparse matrix structure and not the element values themselves.  For matrix-free evaluation, \textsc{Fulqrum} typically uses less peak memory than \textsc{SBD} except as the operator size begins to grow, with Benzene showing an $\sim 2 \rm x$ difference in favor of \textsc{SBD}.  One possible cause of this is that, unlike both \textsc{Dice} and \textsc{SBD}, the operators in \textsc{Fulqrum} are mutable objects that users can transform at the cost of additional copies.  Moreover, the conversion of chemistry problems into qubit Hamiltonians currently requires having two copies of the operator as one must construct the fermionic operator representation.  This added memory is a known issue and, as discussed in Sec.~(\ref{sec:discussion}), there are future optimizations that will reduce this overhead.  This can also be a sign that the caches used in \textsc{Fulqrum} for fermionic systems need further optimization for large molecular sizes. 

\begin{figure}[h]
    \centering
    \includegraphics[width=10cm]{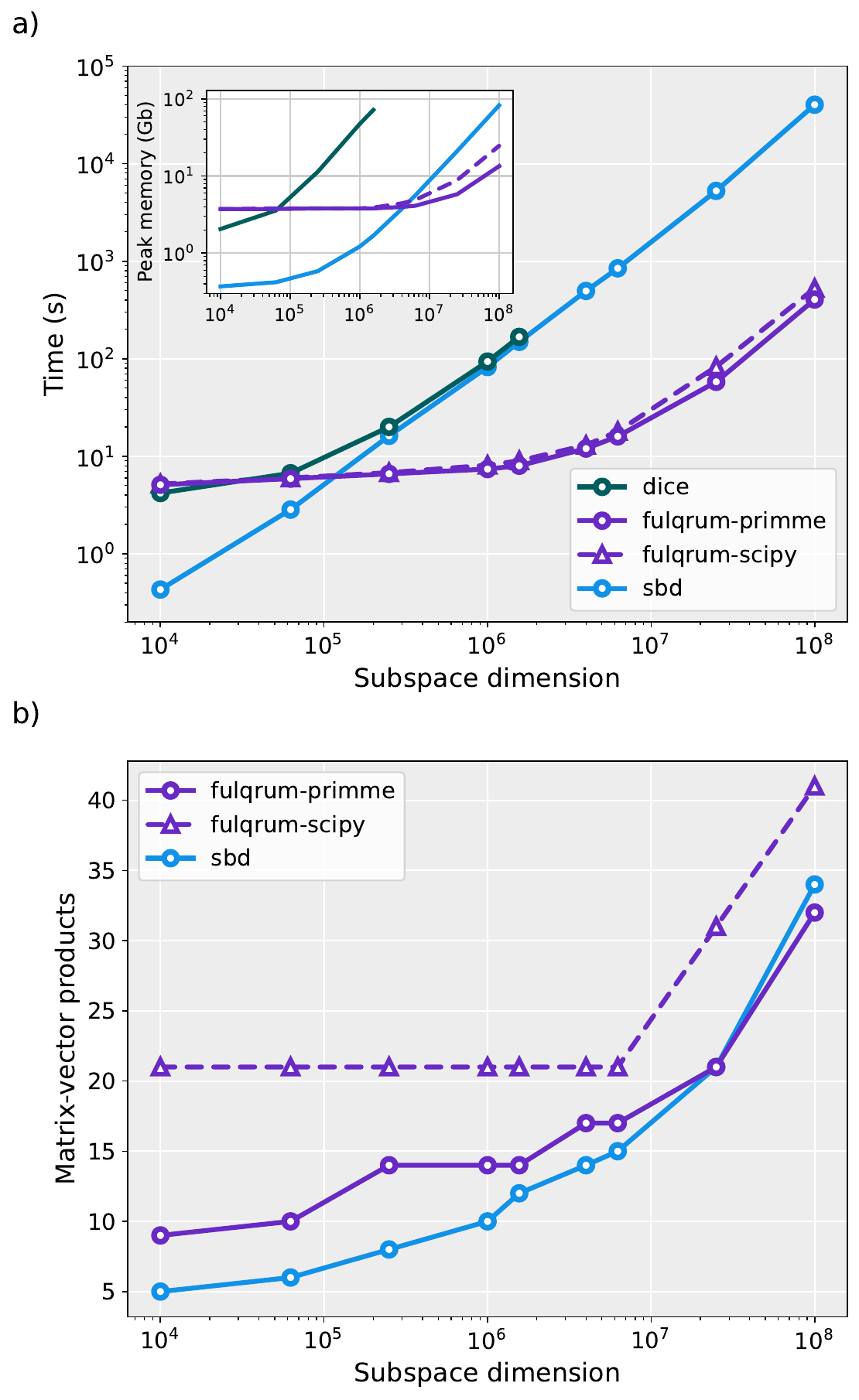}
    \caption{(a) Runtime as a function of subspace dimension for solving the ground state of $\mathrm{Fe}_{4}\mathrm{S}_{4}$ on Computer A.  Solvers are configured such that the relative difference in energy between solutions is $\mathcal{O}(10^{-7})$. Here both \textsc{Fulqrum} and \textsc{SBD} utilize matrix-free evaluation whereas \textsc{DICE} constructs a sparse representation and is thus strictly limited by the available memory of Computer A.  Inset gives the corresponding peak memory usage during the end-to-end eigensolving process. (b) Number of matrix-vector products (SpMV) needed for convergence for the corresponding results presented in (a). Only packages that expose this information are presented.}
    \label{fig:fe4s4}
\end{figure}

In Fig.~(\ref{fig:fe4s4}) we use Computer A, focusing attention on $\rm Fe_{4} \rm S_{4}$ and sweep over different subspace dimensions looking at the scaling trends of runtime, memory usage, and number of SpMV needed to reach convergence. This elucidates some of the performance trade-offs that \textsc{Fulqrum} makes in its design decisions.  At small subspaces dimensions \textsc{Fulqrum} spends time converting an input \text{FCIDump} file into a fermionic operator, and then performing an extended JW transformation to convert this operator into the qubit form needed for eigensolving.  As these steps are unique to \textsc{Fulqrum}, there is an unfavorable runtime compared to the other solvers in the small subspace limit where this time dominates.  In the more pertinent case of large subspace dimensions, this additional runtime is dwarfed by the SpMV time within the eigensolvers, at which point \textsc{Fulqrum} shows a commanding lead over both \textsc{DICE} and \textsc{SBD}.  At the largest subspace dimension, \textsc{Fulqrum} is $100\rm x$ faster and $6\rm x$ more memory efficient than \textsc{SBD}. To show that this advantage is not tied to our choice of eigensolver, we have included using \textsc{Fulqrum} with the \textsc{SciPy} sparse eigensolver based on \textsc{ARPACK} in Fig.~(\ref{fig:fe4s4}) that shows a similar trend.  Here, the \textsc{SciPy} solver uses the same initial state as in \textsc{PRIMME}, but otherwise uses default parameter values.

A similar story plays out for memory consumption relative to the other frameworks.  The subset of Fig.~(\ref{fig:fe4s4}a) indicates that due to \textsc{Fulqrum}'s use of the more expressive operator representation, and the copies required to manipulate and transform these objects, the memory usage is larger than that of others at small dimensions.  However, at larger scales this overhead is again surpassed by the memory of storing the subspace itself and eigensolving, e.g. the number of Krylov vectors before restarting \cite{saad:2011} and/or cached values used for accelerating SpMV.  See Sec.~(\ref{sec:fermi}) for a discussion on caching as it relates to quantum chemistry applications.  The largest subspace in Fig.~(\ref{fig:fe4s4}) reproduces the original SQD experiment, Ref.~\cite{robledo:2024} that took 64-nodes of the Fugaku supercomputer, $3072$ cores and $2$~Tib of memory, to diagonalize the $10^{8}$-dimensional subspace in 90 minutes using \textsc{Dice}.  Here we see that \textsc{Fulqrum} can perform a similar calculation using only $13$~Gb of free-memory in $\sim 7$ minutes; a two- and three-orders of magnitude reduction in both memory consumption and runtime, respectively, whilst running on commodity hardware. 

To isolate whether this performance difference is related to the convergence properties among eigensolvers, or if it is inherent to \textsc{Fulqrum} itself we show the number of SpMV in Fig.~(\ref{fig:fe4s4}b).  At large subspace sizes computing SpMV dominates the total runtime, and therefore the performance is well characterized by the number of SpMV that are performed.  We see that despite having a faster runtime, when using either \textsc{PRIMME} or \textsc{SciPy}, \textsc{Fulqrum} performs \textit{more} SpMV than \textsc{SBD} for all but the largest subspace dimensions.  We can therefore confidently say that the choice of eigensolver is not the determining factor for the performance gains seen in Fig.~(\ref{fig:fe4s4}a).  Rather, it is the difference in the computation time per SpMV that is responsible for the speed up seen in \textsc{Fulqrum}.  This difference can be approximated from Fig.~(\ref{fig:fe4s4}) if we assume that all the runtime is associated with SpMV, a good assumption at larger subspace dimensions, that yields an up to a $93x$ per matrix-vector product runtime advantage for \textsc{Fulqrum} over \textsc{SBD}.  The source of this advantage is three data caching techniques specific to quantum chemistry applications, discussed in Sec.~(\ref{sec:fermi}), that greatly reduce the cost of computing matrix-element values.  Finally we note that Fig.~(\ref{fig:fe4s4}b) shows that the number of SpMV needed for fixed convergence criteria increases with the subspace dimension.  The choice of eigensolution method therefore becomes critical for solving large-scale problems, and there is a crossover point where the \textsc{PRIMME} eigensolver begins to converge faster than that of \textsc{SBD}.  This crossover explains the widening performance gap between matrix-free \textsc{Fulqrum} and \textsc{SBD} on $\rm Fe_{4} \rm S_{4}$ seen in Fig.~(\ref{fig:chemistry-results}).  Therefore, \textsc{Fulqrum}'s flexibility in the choice of eigensolver becomes advantageous when diagonalizing at scale, or when other molecule specific features such as a close to degenerate eigenspectra make the choice of eigensolution method critical.

\section{Discussion}\label{sec:discussion}

We have demonstrated a framework for the solution to eigenproblems for qubit and fermionic systems with performance that is better than other widely used tools in terms of time-to-solution and memory consumption. Our method relies on efficient operator representations and transformations that, along with caching optimizations for the significant problem domain of quantum chemistry, allow for performant SpMV operations either through sparse matrices or via matrix-free evaluation.  This is aided by the flexibility inherent in framework, where users are free to select their classical eigensolver of choice that can further optimize the performance of specific problem instances.  In addition to showing that SQD problems can be faithfully computed with modest computing requirements, our \textsc{Fulqrum} package is easily installed using standard \textsc{Python} packaging tools, further reducing the barrier to entry, as compared to the other tools.  Although we have focused on eigensolving in this work, given that \textsc{Fulqrum} outputs canonical operator formats supporting SpMV, these objects can be utilized in many quantum subspace sparse linear algebra routines. 

As with any piece of software, the performance of the version of \textsc{Fulqrum} presented here can likely be improved upon in several ways. To begin, Message-passing interface (MPI) support should be implemented for CPU execution, and \textsc{Fulqrum} needs to move from a hybrid framework leveraging \textsc{Cython} to a \texttt{C++} centric code base.  The latter would allow for in-depth profiling for bottlenecks.  In addition, as seen in Sec.~(\ref{sec:results}) the memory consumption of \textsc{Fulqrum} should be improved, especially as the size of the input operators grows.  First, it is possible to bypass fermionic operator construction and go directly to a qubit Hamiltonian when importing molecules from \textsc{FCIDump} or \textsc{PySCF} by performing the EJW transformation during import.  Second, although we currently split the qubit operator into diagonal and off-diagonal components before eigensolving, it is possible to compute all the needed values without the implicit data copy.  Together with additional operator and caching optimizations, these modifications should reduce the memory footprint of \textsc{Fulqrum}.  

Finally, and perhaps most importantly, understanding how to optimize workflows that include subspace refinement methods, such as the RAMPS technique presented here, will help to better address the question as to whether SQD applications may offer a viable pathway to near-term Quantum Advantage.  We hope that \textsc{Fulqrum}, or a community-based derivative, can help to answer this in the affirmative.

\section{Methods}\label{sec:methods}

Here we give a detailed description of the implementation of \textsc{Fulqrum}. Our method is primarily written in \texttt{C++}17 with parallelism provided by \textsc{OpenMP} and \textsc{Python} bindings utilizing \textsc{Cython} \cite{cython} that take advantage of both \textsc{NumPy} \cite{numpy} and \textsc{SciPy} \cite{scipy}.  

The canonical \textsc{Fulqrum} workflow is presented in Fig.~(\ref{fig:workflow}) where it is shown that the role of \textsc{Fulqrum} is to take bit-strings sampled from a quantum processor representing the subspace of interest, together with the operator representation of a system Hamiltonian, and produce SpMV-compatible numerical data structures that can be passed to a wide range of eigensolvers.  This procedure can be partitioned into three primary components: Hamiltonian representations, subspace data structure, and matrix evaluation.

\subsection{Hamiltonian representations}\label{sec:hamiltonian}

\subsubsection{Qubit Hamiltonians}\label{sec:qubit}

Our goal is to build numerical representations compatible with SpMV, e.g. Compressed Sparse Row (CSR) matrices or matrix-free methods, from qubit Hamiltonians that are expressed as linear combinations of terms, each comprised of a word of qubit operators $\hat{W}_{i}$ together with a complex-valued coefficient $c_{i}$: $\hat{H} = \sum_{i}c_{i}\hat{W}_{i}$.  Nominally, the alphabet in which words are expressed is the Pauli operators $\hat{I}$, $\hat{Z}$, $\hat{X}$, and $\hat{Y}$, that are valid for execution on a quantum computer.   However, because we perform eigensolving classically, we are free to work with an \textit{extended alphabet} that includes projection operators $\hat{0}$, and $\hat{1}$, together with the ladder operators $\hat{a}^{-}$ and $\hat{a}^{+}$.  In what follows we drop the identity operator as it plays a trivial role in calculations; terms are represented using a sparse encoding.  Internally each term is represented as an \texttt{OperatorTerm} with complex-valued coefficient together with two vectors, one indicating the indices on which individual non-identity operators act, sorted in ascending order, and a vector of characters labeling the operators themselves.  In this work we use little-endian ordering with indexing starting from the right-most bit. Operators are mapped to integers as $\left[Z, 0, 1, X, Y, -, +\right] \rightarrow \left[0, 1, 2, 3, 4, 5, 6\right]$, where off-diagonal operators correspond to values $>2$.  The \texttt{QubitOperator} class stores the vector of terms corresponding to a full Hamiltonian $H$.

Of particular importance for matrix evaluation are the qubit indices on which the off-diagonal operators of a term, if any, act.  We will call this set of indices the \textit{off-diagonal structure} as it plays an out-sized role in efficiently computing matrix-elements.  Restricting ourselves to Pauli-words, the following hold true for terms in a Hamiltonian described over a subspace of the computational basis represented by bit-strings:

\begin{enumerate}
  \item\label{item:one_element} For a given bit-string from the subspace representing a row in the Hamiltonian matrix, there exists exactly one nonzero element per term; The matrix representation of a term in the subspace has a single nonzero entry per row \footnote{The same holds true for the columns due to the Hermitian structure of Hamiltonians}. For example, the operator $XXY$ has non-zero values along the anti-diagonal.
  \item\label{item:column} The bit-string for the column corresponding to the nonzero element from (\ref{item:one_element}) is found by flipping the bits of the row bit-string at the off-diagonal structure indices; the off-diagonal indices give a bit-mask that XOR with the row bit-string to yield the column bit-string. e.g. given a row bit-string $011$ and operator $XZY$, the column bit-string for the corresponding nonzero element is $110$.
  \item\label{item:groups} As a corollary to (\ref{item:column}), terms that share the same off-diagonal structure correspond to the same matrix-elements\footnote{\textsc{PennyLane} \cite{pennylane} uses similar reasoning to group Pauli-terms, but their grouping requires forming a matrix with dimension equal to the product of the number of Pauli-terms and the operator width; a matrix that can exceed gigabyte memory sizes for problems such as those from quantum chemistry.}.  As an example, the terms $(XZX, YIX, YZX)$ have the same off-diagonal structure and therefore correspond to the same matrix elements, where as $(ZXY, IYX, ZXX)$ have a different off-diagonal structure and therefore contribute to different matrix elements.
\end{enumerate}

\begin{figure}[t]
\centering
\includegraphics[width=8cm]{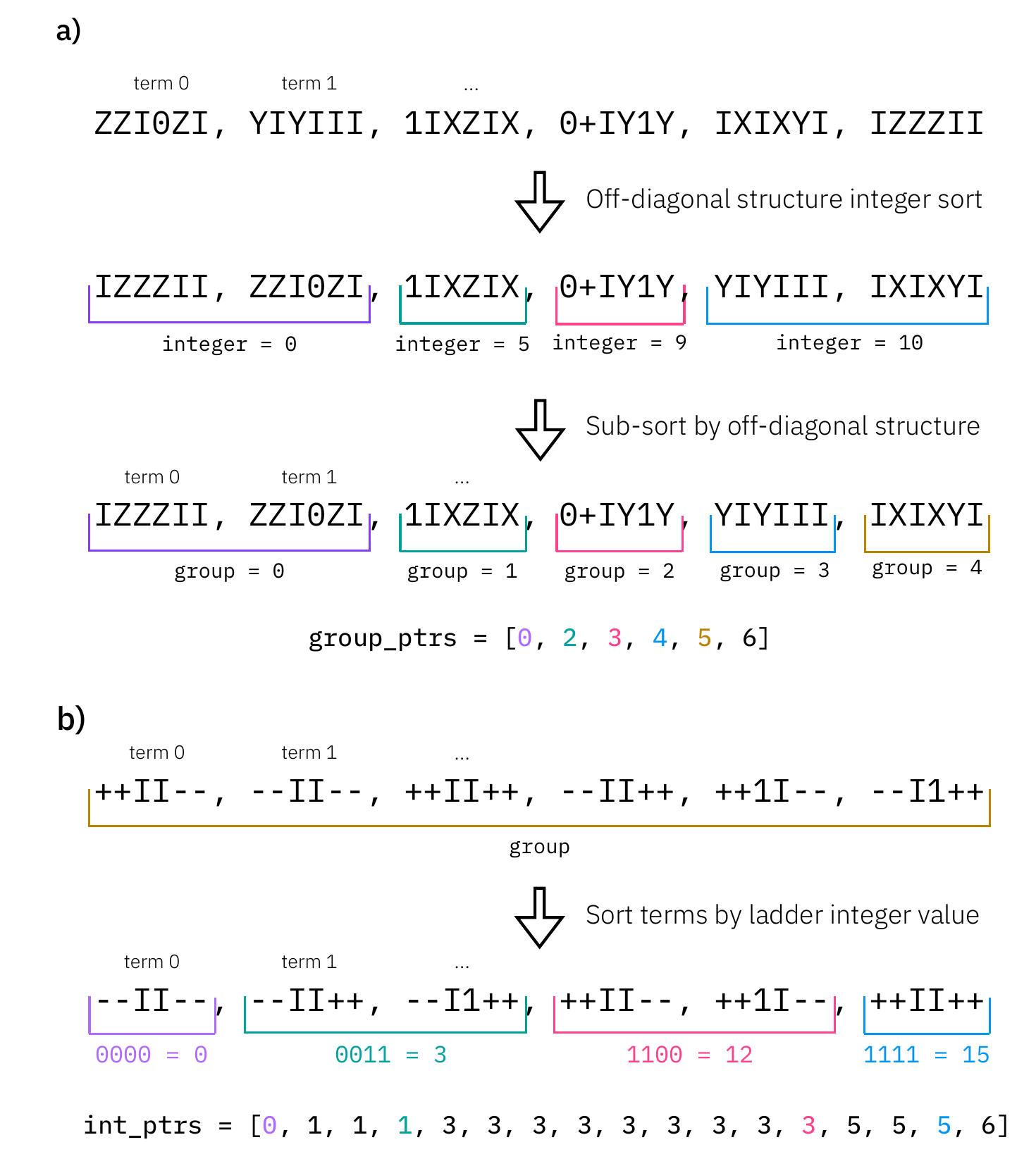}
  \caption{Term sorting methods used in \textsc{Fulqrum}. a) Groups of terms corresponding to matrix elements are sorted into contiguous order by first ordering all terms by the integer sum of their off-diagonal indices, and then sub-sorting terms with matching integer values by their off-diagonal structure.  Both steps can be done in parallel.  The starting and stopping term indices for each group are stored in the group pointers array. b) Sorting by ladder integer value within a group comprised from terms with off-diagonal structure determined by ladder operators.}\label{fig:groupings}
\end{figure}

Terms expressed over the full extended alphabet follow the same set of rules with the caveat that the single matrix-element per row can be zero if the term is comprised of one or more projector and/or ladder operators.  Such operators place constraints on the bit-values at the corresponding operator indices for which nonzero values occur.  For a word with $m$ projection or ladder operators, the fraction of rows with nonzero entries is reduced to $1/2^{m}$.  Although qubit Hamiltonians are nominally not expressed using these extended operators, as discussed in the subsequent section, fermionic systems naturally utilize these operators, and we can leverage these constraints to greatly reduce the number of terms that need to be computed for matrix evaluation.

\subsubsection{Fermionic Hamiltonians}
Fermionic systems differ from those of qubit Hamiltonians in two important ways.  First is the use of fermionic ladder operators, $f^{\dag}$ and $f$, that obey the canonical anti-commutation relations: $\{ f_{i}, f_{j}\} = 0$ and $\{f_{i}, f^{\dag}_{j}\} = \delta_{ij}$. And second is the possibility of having multiple operators act on the same mode index. Despite these differences, it is beneficial to represent fermionic terms (\texttt{FermionicTerm}) and Hamiltonians (\texttt{FermionicOperator}) using the same data structures as qubit Hamiltonians. Likewise, we sort indices in ascending order, keeping track of sign changes when exchanging operators on different indices.  In addition, we merge operators acting on repeated indices within a term according to the rules in Tbl.~(\ref{tab:merge}), which yields an operator comprised of both ladder operators and projectors.  While it is possible to directly build fermionic operators in \textsc{Fulqrum}, standard workflows involve converting fermionic operators from \textsc{FCIDump} files or tools such as \textsc{PySCF}, \textsc{OpenFermion} \cite{openfermion}, or \textsc{Qiskit Nature} \cite{qiskitnature} into their equivalent \textsc{Fulqrum} representations. In particular, we have optimized conversion from \textsc{FCIDump} and \textsc{PySCF} so that only unique fermionic terms are present in the final \texttt{FermionicOperator}; collecting repeated terms need not be performed when importing from these formats.

\begin{table}[h]
	\begin{tabular}{ll}
		\\
		\hline
		$-- = ff = \rm null$ & $+- = f^{+}f \equiv 1$\\
		$0- = ff^{+}f = f(1-ff^{+}) = -$ & $1- = f^{+}ff = \rm null$\\
		$-+ = ff^{+} \equiv 0$ & $++ = f^{+}f^{+} = \rm null$ \\
		$0+ = ff^{+}f^{+} = \rm null$ & $1+ = f^{+}ff^{+}= f^{+}(1-f^{+}f) = +$ \\
		$-0 = fff^{+}=\rm null$ & $+0 = f^{+}ff^{+} = f^{+}(1-f^{+}f) = +$ \\
		$00 = ff^{+}ff^{+} =ff^{+}(1-f^{+}f) = 0$ & $10 = f^{+}fff^{+}= \rm null$ \\
		$-1 = ff^{+}f = f(1-ff^{+})= -$ & $+1 = f^{+}f^{+}f = \rm null$ \\
		$01 = ff^{+}f^{+}f = \rm null$ & $11 = f^{+}ff^{+} = f^{+}f(1-ff^{+})= 1$\\
		\hline
	\end{tabular}
    \caption{Complete table of merging rules for combining repeated indices in fermionic terms over the alphabet of projection and ladder operators where $0\equiv ff^{+}, 1\equiv f^{+}f$. `null' indicates those outcomes that do not contribute to the output operator.}\label{tab:merge}
\end{table}

As we aim to have a unified solution method, we transform fermionic operators into qubit operators via an extended Jordan-Wigner (EJW) transformation \cite{jordan:1928}.  In contrast to the standard JW transformation, we do not decompose fermionic ladder operators into Pauli operators, $f_{j}=\frac{1}{2}\left(X_{j}+i P_{j}\right)Z_{j-1}\otimes \dots \otimes Z_{0}$, and instead directly map projection and ladder operators into their qubit counterparts along with adding the $Z$ operators needed to maintain commutation relations. JW transformations are often used, for example, in mapping fermionic problems for variational algorithms \cite{cerezo:2021}. However, these transformations can bottleneck a computation before quantum is evoked.  As shown in Tbl.~(\ref{table:jordanwigner}), the total time of the EJW transformation presented here is distinctly faster than that of the corresponding JW transformation utilized in common frameworks.  With the EJW transformation time much smaller than the eigensolving runtime for fermionic systems, requiring this transformation in order to have a unified solution method does not represent a bottleneck in our method.

\begin{table}[b]
\begin{tabular}{l|c|c}
Framework & Time (sec) & Memory (Gb)\\
\hline
\textsc{cudaq-algorithms} (0.1.0) & $2,394$ & $92.4$\\
\textsc{ffsim} (0.0.84) & - & OOM \\
\textbf{\textsc{fulqrum}} & $\mathbf{6}$ & $\mathbf{3.9}$\\
\textsc{openfermion} (1.8.1)& $3,913$ & $26.6$\\
\textsc{pennylane} (0.45.1) & $86,400+$ & -\\
\textsc{pychemiq} (1.1.4) & - & OOM\\
\textsc{qiskit-fermions} (0.3.0) & - & OOM\\
\textsc{qiskit-nature} (0.7.2) & - & OOM\\
\textsc{qrisp} (0.9.6)& $36,252$ & $34.7$\\
\end{tabular}
\caption{Time and memory taken by Computer A to compute the EJW transformation for Benzene in the 6-31G basis generated by \textsc{PySCF} from this work along with a selection of software packages performing a standard JW transformation\cite{cudaqalgs, ffsim, openfermion, pennylane, pychemiq, qiskitfermions, qiskitnature, qrisp}.  The fermionic operator contains $\sim 44.2$ million total terms over $132$ operator indices.  $86,400+$ indicates times that took longer than 24 hours, whereas Out Of Memory (OOM) signifies that the routine used more memory than Computer A has available.}
\label{table:jordanwigner}
\end{table}

At the output of the EJW transformation the resulting qubit Hamiltonian is expressed with operators in the set $\{ Z,0,1,-,+\}$, with the off-diagonal structure determined solely by ladder operators.   As stated in Sec.~(\ref{sec:qubit}), these extended operators constrain the possible bit-values for which a row bit-string yields a nonzero value for a given term. We can leverage this on a group by group basis by converting the ladder operators in each term to the binary representation of an integer with the mapping $-\rightarrow 0,+\rightarrow 1$; a value we call the terms \textit{ladder integer}.  The terms within the group are then sorted in ascending order according to these ladder integer values.  Associated with each group is an array of pointer values (\texttt{int\_ptrs}) that records the starting and stopping indices, with respect to the full Hamiltonian, of terms within the group that have a given ladder integer value.  Molecular electronic Hamiltonians can be written in the form
\begin{equation*}\label{eq:fermionic}
H = H_{0} + \sum_{p,q}h_{p,q}f_{p}^{+}f_{q} + \frac{1}{2}\sum_{p,q,r,s}h_{p,q,r,s}f_{p}^{+}f_{q}^{+}f_{r}f_{s}
\end{equation*}
where $H_{0}$ is a constant offset nuclear repulsion energy and $h_{p,q}$ and $h_{p,q,r,s}$, are one- and two-body integrals, respectively \cite{helgaker:2000}, showing that at most four ladder operators are present in a term.  We therefore need a pointer array of size $2^{4}+1$ to accommodate all possible ladder integer values. An example is given in Fig.~(\ref{fig:groupings}b).  For each off-diagonal group in the Hamiltonian we compute the ladder integer for a given row bit-string using the indices comprising the off-diagonal structure for the group.  Only those terms in the group that have a matching ladder integer value give a nonzero contribution to the resultant matrix-element,  assuming the column bit-string is also in the subspace. While this reduces the number of terms that need to be numerically evaluated, the primary savings come from those groups where no term shares the row bit-string ladder integer value; there is no column lookup, or any other processing, for the group as the matrix-element is zero.

\subsection{Subspace data structure}\label{sec:subspace}

The \texttt{Subspace} in \textsc{Fulqrum} is a collection of bit-sets \cite{boost}, \texttt{boost::dynamic\_bitset<size\_t>}, defining the rows and columns of a Hamiltonian matrix. \textsc{Fulqrum}'s algorithm requires both insertion order indexing and fast lookup operations, and therefore the choice of subspace data structure is paramount for performance and memory-efficiency. A key routine in \textsc{Fulqrum} is fetching a row bit-set by index and constructing a potential column bit-set by flipping bits at the indices specified by the off-diagonal structure in the row bit-set. We then lookup the candidate column bit-set in the subspace. If the bit-set is in the subspace, the matrix element between that row and column needs to be evaluated. Because we filter out nonzero values \textit{before} querying the subspace, this element value is nonzero, and should be included in the output matrix, or computed for matrix-free SpMV. To ensure sufficient performance, we need a data-structure that allows for fetching multiple (row) bit-sets in parallel, and that also offers fast search capabilities. Here we use \texttt{emhash8::HashMap}~\cite{emhash} to store and look up bit-sets as it provides both features, and \texttt{rapidhash}~\cite{rapidhash} as the bit-set hasher.

Most \texttt{C++} hash maps, including \texttt{std::unordered\_map}, do not allow parallel iteration over keys. Had we chosen a different hash map, we would need a vector to store bit-sets and to allow for parallel iteration, whilst also maintaining the hash map for fast look up. Having two data structures to store identical information would necessarily increase the memory footprint of \textsc{Fulqrum}.  \texttt{emhash8::HashMap} allows for indexed-access to key-value pairs in insertion order, similar to a \texttt{std::vector}.  We can therefore parallelize the access to keys using, e.g., using \texttt{OpenMP}, thus achieving both parallelism and fast look up in a single data structure.

\texttt{emhash8::HashMap} stores key-value pairs in \emph{buckets}. During a key look up, it follows a standard hash map look up routine where the key is first hashed and then a hash value, modulo the number of buckets, is performed to find a potential bucket index where the key might be stored. If the bucket at that index is \emph{empty}, the key is guaranteed not to be in the hash map. To improve search performance, we edited the \texttt{emhash8::HashMap} implementation to include a variable to store the bucket occupancy information. The variable is a \texttt{boost::dynamic\_bitset<size\_t>} with a width (number of bits) equal to the number of buckets. Initially, all bits are off in this bit-set , indicating empty buckets. When we put a key-value pair in a bucket during insertion, the corresponding bit will be set to on. During search, we first inspect this bucket occupancy bit-set to validate the bucket is occupied before looking into the main bucket structure. If the bit is off, the bucket is empty and the search terminates yielding a faster way to reject empty buckets. For example, for a 64 bytes cache-line load, the bit-set-based implementation can bring occupancy information about $64 \times 8 = 512$ bits (buckets) in the cache, improving the cache hit rate. Had we directly queried main bucket structure, which has at least two \texttt{size\_t} ($2 \times 8$ bytes) items per bucket, a single cache-line load, theoretically, only brings $64 / (2 \times 8) = 4$ buckets in the cache.  Note that we can leverage this faster empty bucket test because we know the number of key-value pairs in our hash map a priori. This allows for reserving storage space and fixing the total number of buckets in the hash map, ultimately informing us of the number of required bits in the bit-set. Moreover, we can reserve more capacity in the hash map than necessary to keep the hash map load factor low. A lower load factor improves search performance, especially for missing items, as we can reject empty buckets faster using our modification. 

Subspaces can be further optimized for domain specific applications.  For example, in quantum chemistry, a fermionic case, it is often beneficial to view each bit-string (determinant) as the product of spin-up [alpha (A)] and spin-down [beta (B)] orbitals.  For a bit-string of length $2N$, this is represented in \textsc{Fulqrum} as $|\beta_{(N-1)}\dots \beta_{1}\beta_{0}\alpha_{(N-1)}\dots \alpha_{1}\alpha_{0}\rangle$, and the \texttt{Subspace} object is constructed from two lists, one for the alpha half bit-strings and one for the beta half bit-strings. A tensor product subspace (TPS) can be constructed by taking the tensor product of the alpha and beta half bit-strings.  For alpha and beta lists of dimension $d_{A}$ and $d_{B}$, respectively, the total TPS dimension is  $d = d_{A}\cdot d_{B}$, resulting in large subspace dimensions for even modestly sized inputs.  Moreover, additional tensor product bit-strings are unlikely to be in the resultant eigenvector, even if the bit-strings used to generate alpha and beta lists are; TPS are brute-force methods for generating physically relevant bit-strings.   However, TPS have computational benefit in that it is possible to completely bypass the costly subspace lookup operations.  As these are the dominant bottleneck in the SpMV calculations, using TPS can dramatically accelerate the eigensolving process.  This is particularly true for matrix-free methods where many repeated subspace lookup operations would otherwise have to be performed.  



\subsection{Matrix evaluation}\label{sec:evaluation}

\subsubsection{General method}

Numerical representations of a subspace Hamiltonian are generated by first combining a \texttt{Subspace}, $\mathcal{S}$, and \texttt{QubitOperator} into a single \texttt{SubspaceHamiltonian} class. Within this object the Hamiltonian is split into diagonal and off-diagonal components, with the diagonal matrix elements within the subspace being computed in parallel and cached.  This calculation is accelerated when working over terms with projection operators, such as those for fermionic systems, where it is possible to efficiently pass-over null terms based on bit-values of the row bit-string at the projector operator indices. Off-diagonal Hamiltonian terms are sorted into groups, and if the Hamiltonian is derived from a fermionic system, groups are further sorted by their terms ladder operator (off-diagonal structure) derived integer values.

The primary output format for a subspace Hamiltonian is a CSR sparse matrix, selected for its wide-compatibility with classical eigensolving frameworks.  It is possible to build a CSR matrix directly, with no intermediate copies, using a two-pass method where the first pass computes the number of nonzero entries per row of the Hamiltonian (the sparse structure) while the second pass populates the matrix element values.  Both stages proceed by independently iterating over the bit-strings of a subspace in parallel, with each bit-string representing a row of the CSR matrix.  The cached diagonal elements are processed first, while the off-diagonal Hamiltonian is treated group-by-group.  For a given row-bit-string we only evaluate those groups with matrix-elements in the lower-triangle.  If the Hamiltonian is derived from a fermionic system then the ladder integer value for the bit-string determined by the off-diagonal structure for the group is computed and used to index the \texttt{int\_ptrs} corresponding to the group which sets the start and stop indices used to lop over nonzero terms in the group.  If there are no matching ladder integer terms in the group then the start and stop indices are identical and the group is skipped. 

For groups with nonzero terms, we first compute the corresponding column bit-string for the matrix-element using the off-diagonal structure, and perform a hash map lookup to determine if this column bit-string is within the subspace.  If so, then the relevant terms within the subspace, i.e. all terms for spin Hamiltonians and those with ladder integer values that match the row bit-string for fermionic systems, are further refined by considering the value of bits at the location of projection operators, if any, to further identify zero terms.  At the end of this process, only terms with nonzero contribution to the matrix-element are numerically evaluated and summed. 

If computing the sparse structure (first pass), the number of nonzero elements per matrix row are stored in array, \texttt{indptr}, of length $\dim(\mathcal{S})+1$, with the last element representing the total number of nonzero elements (\texttt{NNZ}) in the matrix.  The second pass uses structure information from pass one to construct arrays of column indices (\texttt{indices}) and matrix-element values (\texttt{data}), both of length \texttt{NNZ}, and populates them by processing the matrix again.  The trinity (\texttt{indptr}, \texttt{indices}, \texttt{data}) is used to build a \textsc{SciPy} \texttt{csr\_matrix} that is then wrapped in a \texttt{CSRLinearOperator} that allows overloading the CSR SpMV operation to allow for \textsc{OpenMP} parallelization.  \textsc{Fulqrum} automatically uses the smallest data-types for both the matrix data (\texttt{double} or \texttt{double complex}) and row and column indices (\texttt{int32} or \texttt{int64}) both internally and for the output CSR matrix.   

While the two-pass method allows for direct CSR construction with no copies, it is often computationally expensive to evaluate matrix-elements twice.  As such, we introduce a ``fast'' CSR construction method that utilizes an array of $\dim(\mathcal{S})$ objects comprised of dynamically allocated \texttt{C++} vectors for column indices and matrix-element values to populate a data structure similar to CSR format in a single step.

In situations where memory constraints make explicit matrix storage impractical, it is possible to perform the SpMV operations required for sparse eigensolving using matrix-free methods. However, these  techniques do not alleviate the memory requirements for storing the subspace and Hamiltonian.  In \textsc{Fulqrum}, the \texttt{SubspaceHamiltian} is a subclass of the \textsc{SciPy} \texttt{LinearOperator} class and, replacing a matrix, can be directly passed to eigensolvers supporting this format for matrix-free evaluation, e.g. \cite{scipy, primme,slepc}. Given in input vector, SpMV computation is performed in parallel over row bit-strings in a similar manner to CSR matrix construction, where the action of each row bit-string on the input vector is used to populate the value of the output vector at same row index. Because no data are stored between runs, each SpMV call requires recomputing matrix-elements; each matrix-free SpMV call is comparable to fast CSR matrix construction in terms of runtime.  As such, the number of iterations performed by the eigensolver dictates the overall runtime of our matrix-free algorithm.  It is therefore most useful when teamed with a good initial trial vector to accelerate convergence, or when no other method is feasible due to memory constraints. However, we can use caches to accelerate repeated matrix element computation in Fermionic cases (Sec.~\ref{sec:fermi}).

As we are working with Hermitian matrices, it is possible to completely characterize the Hamiltonian matrix using matrix-elements in the the upper or lower triangle only. We can exploit this opportunity to cut down the number of matrix element evaluations by half, possibly speeding up CSR matrix build time by $\sim 2\rm x$. While this is conceptually straightforward, there are implementation challenges in direct implementation. Importantly, we need an efficient way to determine if a matrix-element is in the upper or lower triangle. Focusing on the lower-triangle, for an element to be in the lower triangle, the index of a row bit-string must to be greater than the index for the associated column bit-string for a given group. Although we can compute the column bit-set first and perform a column lookup, the cost of doing so offsets any gains from looking at the lower or upper triangle only.  To get around this, we first sort subspace bit-strings by their integer values.  This takes only a small fraction of time with respect to the overall eigensolution runtime, and can be performed in parallel.  We can then determine if the matrix-element for a given group is within the lower triangle by inspecting the row bit at largest index in the groups off-diagonal structure; the \textit{most-significant off-diagonal bit} (MSOB). If the bit at the MSOB position of a row bit-string is on, then it will be flipped to off when computing the column bit-string, and the integer sorting of the subspace bit-strings guarantees that the matrix-element is in the lower-triangle of the Hamiltonian.  Similarly, if the MSOB row bit is off then the corresponding bit in the column bit-set will be on, and we know the element is in the upper-triangle.  In this way, we are able to compute only lower/upper triangle matrix-elements without computing the column bit-string itself.

\subsubsection{Fermionic optimizations}\label{sec:fermi}

Solutions to fermionic systems can often be accelerated by taking advantage of physical properties and symmetries of the underlying Hamiltonian and subspace.  As optimizations for fermionic subspaces used in quantum chemistry, e.g. TPS, was discussed in Sec.~(\ref{sec:subspace}), here we focus on properties of the Hamiltonian.  For systems from quantum chemistry, Eq.~(\ref{eq:fermionic}) indicates that only off-diagonal terms with two (single excitation) or four (double excitation) ladder operators exist; row and column bit-strings are connected by two or four bit-flips. The groups of the Hamiltonian can then be classified by which spin-sector the ladder operators reside in.  AA / BB - single excitation terms with both flips in the same spin-sector, alpha and beta respectively, AABB - double excitations with two ladder operators in both alpha and beta halves, and finally AAAA / BBBB - all four flips occur in the alpha or beta sector alone.  Other combinations are not possible as the number of set bits in each sector should be fixed.
With this identification in place, it is possible to construct efficient caching methods that allow for data reuse and reduce the number of loops required to compute matrix element values.

The caching techniques used by \textsc{Fulqrum} for fermionic system are derived from the following:

\begin{enumerate}
  \item\label{item:aaaa} For AAAA and BBBB groups, all literals, including diagonal operators, $(Z,0,1,-,+)$ occur only in either the alpha- or beta-sector, respectively; matrix-elements between row and column bit-strings are determined by a single spin-sector alone.
  \item\label{item:aa} AA / BB groups have ladder operators in a single spin-sector, and there is at most a single diagonal literal in the opposite sector, e.g. $(Z, 0, 1)$.
  \item\label{item:aabb} In AABB groups, all terms have the same coefficient, and only a single term contributes to the matrix-element; only a single term satisfies the ladder integer criteria.
\end{enumerate}

For AAAA / BBBB groups, item (\ref{item:aaaa}) allows us to go through every possible set of alpha (or beta) pairs that are a Hamming distance four apart and compute the matrix element for each selected pair. We then cache these values for both permutations of the pair. When building the CSR matrix or performing matrix-free evaluations we use this cache as a lookup, and are able to skip repeated for-loops needed for matrix-element evaluations.  While the full list of alpha bit-strings can be large, Slater-Condon rules restrict valid pairs to those bit-strings that are a Hamming distance four apart.  This greatly limits the permissible number of pairs and the resulting size of cached values.  

We could cache values for AA / BB groups in a similar fashion if it were not for the possibility of a single literal on the opposite spin-sector, item (\ref{item:aa}).  For a single AA group and an alpha pair $(\alpha_{0}, \alpha_{1})$ we can combine the contribution of those terms that have no literal value in beta into a single value $c_{0}$.  The beta-dependent terms can be stored as an array of values for each $(\alpha_{0}, \alpha_{1})$ pair that we call $\texttt{cpb}[k]$ (cache-per-beta), where $k$ indexes the beta component.  The single literal indicates that there are two possible values for each beta ($k$) in $\textrm{cpb}$ depending on if the corresponding bit is set or not, and a naive implementation would need to store both values.  However, we can reduce the storage space by a factor of two by adding the beta contributions when the bit is not set to $c_{0}$, i.e. $c'_{0}=c_{0}+\sum_{k}\texttt{cpb}[k][\textrm{bit}=0]$, and computing the beta dependent part as the difference between set and unset bit values $\delta[k] = \texttt{cpb}[k][\textrm{bit}=1] - \texttt{cpb}[k][\textrm{bit}=0]$.  Matrix elements can then be computing using a single for-loop $\textrm{element} = c'_{0} + \sum_{k}\textrm{bit}\cdot\delta[k]$ where the loop iteration is bounded by the number of beta orbitals.
However, the loop over beta orbitals can be removed entirely. For an AA group the beta-dependent terms coefficients depend only on the off-diagonal flip indices or alpha orbital pair $(p, q)$.  Hence, up to the fermionic sign $s$, $\delta[k]$ is identical for every alpha pair that shares the same $(p, q)$. Moreover, because the beta half-strings are common to all rows, the sum over beta orbitals can then be precomputed once per beta string $\beta$. It allows us to compute a pair table $T[\beta][pq]$ such that matrix elements reduce to a single lookup, $\textrm{element} = c'_{0} + s \times T[\beta][pq]$. The storage for this approach scales with the number of beta half-strings times the number of alpha orbital pairs, independent of the number of alpha pairs. BB groups are handled in a similar manner.

Finally, for those double-excitation groups with an excitation in each spin sector, AABB, item (\ref{item:aabb}) tells us that only a single-term contributes to the final matrix element value.  Moreover, this value depends on the product of two independent sign terms, one for alpha and one for beta.  We can therefore precompute these values for all valid alpha and beta pairs that are a distance two apart.  The values themselves are determined by the parity of each spin sector, where the parity is equal to the number of set bits between the bit-flip operators in that sector.  Computing matrix-elements then becomes the product of the term coefficient and two efficient cache lookups.  As with other optimizations, we are able to reduce the number of loops needed to compute element values, and better optimize memory access.

Note that the operator caching methods discussed here differ fundamentally from those used in packages such as \textsc{SBD} where bit-strings are cached for repeated use \cite{walkup:2026,doi:2026}.  As the dimensionality of the subspace can be made much larger than the operator size for a molecule, our caching techniques are more memory efficient in the target limit of substantial subspace sizes.

\section{Recursive algorithm for a minimal perturbative subspace}\label{sec:ramps}

Here we describe the method used to generate the synthetic subspaces used for $\rm Fe_{4} \rm S_{4}$ and Benzene.  Our technique takes advantage of the fact that many off-diagonal matrix elements, in problems such as those from quantum chemistry, may be markedly smaller in amplitude than the diagonal energy splittings between states, allowing for their contributions to the final eigenenergy to be treated perturbatively. In addition, we utilize the structure of the Hamiltonian, noting that off-diagonal matrix-elements not directly coupled to an initial target subspace are mediated by additional small couplings, the product of which quickly goes to zero.  The end result is a Recursive Algorithm that generates a Minimal Perturbative Subspace (RAMPS) around an initial subspace. 

To elucidate this method we consider the tri-diagonal Hamiltonian eigenvalue equation:

\begin{align}
\begin{pmatrix}
H_{00} & H_{01} & 0 & 0 \\
H_{10} & H_{11} & H_{12} & 0 \\
0 & H_{21} & H_{22} & H_{23} \\
0 & 0 & H_{32} & H_{33}
\end{pmatrix} 
\begin{pmatrix}
c_{0} \\
c_{1} \\
c_{2} \\
c_{3} 
\end{pmatrix} = E
\begin{pmatrix}
c_{0} \\
c_{1} \\
c_{2} \\
c_{3} 
\end{pmatrix},
\end{align}
where we assume that the $0^{\rm th}$-row contains the dominant contribution to the eigenenergy, $E/H_{00} = 1 +\delta/H_{00}$, where $\delta/H_{00}$ is a small parameter, and that the off-diagonal terms are smaller than the diagonal energy splittings; we start with an initial subspace consisting of a single bit-string and target energy $H_{00}$.  Our task is to determine the energy shift $\delta/H_{00}$ and the form that the interaction takes.

Assuming that the off-diagonal terms are small compared to the energy splittings with the target energy, $H_{00}$, we arrive at a recursive expression for $\delta / H_{00}$:

\begin{align*}
\frac{\delta}{H_{00}} &\simeq \frac{\left|H_{01}\right|^{2}}{H_{00}\left(H_{00}-H_{11}\right)} \stepcounter{equation}\tag{\theequation}\label{eq:ramps}\\
&+ \frac{1}{H_{00}\left(H_{00}-H_{11}\right)}\left[\frac{\left|H_{01}\right|^{2}}{\left(H_{00}-H_{11}\right)} \frac{\left|H_{12}\right|^{2}}{\left(H_{00}-H_{22}\right)}\right] \\
&+ \frac{1}{H_{00}\left(H_{00}-H_{11}\right)\left(H_{00}-H_{22}\right)} \\ 
&\times \left[\frac{\left|H_{01}\right|^{2}}{\left(H_{00}-H_{11}\right)} \frac{\left|H_{12}\right|^{2}}{\left(H_{00}-H_{22}\right)}\frac{\left|H_{23}\right|^{2}}{\left(H_{00}-H_{33}\right)}\right]
\end{align*}

The first line in Eq.~(\ref{eq:ramps}) yields the shift in energy from terms directly coupled to the target subspace.  The remaining lines show the recursive nature of the coupling between bit-strings not directly coupled to the initial subspace, with this interaction mitigated by an increasing number of small off-diagonal matrix-elements leading to imperceptible impact on the final energy.

\begin{algorithm}[t]
    \footnotesize
    \SetKwProg{subproc}{Function}{}{}
    \DontPrintSemicolon
    \caption{RAMPS} \label{alg:ramps}
    \KwIn{An Hamiltonian $H$ partitioned into a set of off-diagonal groups $\mathcal{G}$, initial subspace $\mathcal{S}_{0}$,  target energy $E$, and tolerance $\tau$.}
    \KwOut{Modified subspace $\mathcal{S}_{\rm out}$}

    \subproc{\textnormal{\textsc{RAMPS}(H, $\mathcal{S}_{0}, E, \tau$)}}{
    	$\mathcal{S}_{\mathrm{out}} \gets \mathcal{S}_{0}$ \; 
        \ForEach{$b_{i} \in \mathcal{S}_{0}$}{
          $nxt\_rows \gets \left\{ b_{i}\right\}$ \tcp*{Next rows}
          $nxt\_pre \gets \left[1/E\right]$ \tcp*{Next prefactors}
          	\While{$\mathrm{dim}(nxt\_rows) \neq 0$}
          	{
           		$cur\_rows \gets nxt\_rows$ \tcp*{Current rows}
          		$cur\_pre \gets nxt\_pre$ \tcp*{Current prefactors}
          		$nxt\_rows \gets \{ \}$\;
          		$nxt\_pre \gets []$\;
          		$z \gets 0$ \tcp*{Counter for prefactor array}
           		 \ForEach{$b_{j} \in cur\_rows$}
           		 {
           		 	\ForEach{$g \in \mathcal{G}$}
           		 	{
           		 		$b_{k} \gets $ \textsc{ColumnBitstring}$(j, g)$\;
           		 		$amp \gets cur\_pre [z] \cdot\frac{|H_{jk}|^{2}}{(E-H_{kk})}$\;
           		 		\If{$|amp| > \tau$}
           		 		{
           		 			\tcp*[l]{Add column and prefactor}
           		 			$nxt\_rows \gets nxt\_rows \cup \{b_{k}\}$\;
           		 			$nxt\_pre$.add($amp/(E-H_{kk})$)\;
           		 		}
           		 	}
           		 	$z \gets z + 1$\;
           		 }
           		 $\mathcal{S}_{\mathrm{out}}\gets \mathcal{S}_{\mathrm{out}} \cup nxt\_rows$\tcp*{Add next rows}
       		 }
        }
        \Return{$\mathcal{S}_{\rm out}$}
    }
\end{algorithm}

The expansion in Eq.~(\ref{eq:ramps}) can be generalized to work with arbitrary Hamiltonians $H$ and input parameters.  The simplest scenario, and the one used in this work, is to start with an initial subspace $\mathcal{S}_{0}$, comprised of a single bit-string, and target energy $E$ that is the diagonal energy corresponding to this starting bit-string.  However, for Hamiltonians where the target eigenvalue is nearly degenerate and/or the off-diagonal terms cannot be treated perturbatively, i.e. the system is strongly-coupled, $\mathcal{S}_{0}$ can be a multi-dimensional subspace and the target energy $E$ would be an eigenvalue found by solving $H$ within $\mathcal{S}_{0}$.  For example, if we aim to find the ground state eigenenergy, then the target energy $E$ would be the ground state eigenenergy for $H$ restricted to $\mathcal{S}_{0}$. Together with an user supplied tolerance value $\tau$, these values can be used as the starting point for RAMPS recursion.  The general algorithm is presented in Alg.~(\ref{alg:ramps}).  

Algorithm~(\ref{alg:ramps}) yields a refined output subspace $\mathcal{S}_{\rm out}$ that is comprised of the initial subspace $\mathcal{S}_{0}$ together with bit-strings that correspond to matrix elements that shift the target energy value by a relative amount greater than $\tau$.  The output subspace is therefore nominally larger than the initial. In practice, this algorithm is modified to terminate the RAMPS process after a maximum number of recursions to prevent the possibility of traversing the full connected component of Hilbert space in which $\mathcal{S}_{0}$ resides.  Because the initial subspace and target energy are necessarily approximations (otherwise the solution would be known), it is possible to use the RAMPS output to generate better a trial subspace $\mathcal{S}_{0}$ and associated target energy $E$ for a subsequent RAMPS refinement.  In this manner, it is possible to iteratively yield subspaces with increasing overlap with the true eigenstate, and thus more accurate eigenenergies. 

\bmhead{Acknowledgments}

The authors thank Jim Garrison for helpful discussions and Danil Kaliakin, Zhen Li, and Kenneth Merz for permission to use $\mathrm{CH}_{4}$ dimer bit-strings and operator representation from Ref.~\cite{kaliakin:2025}.

\bmhead{Author contributions}

PDN conceived the project, derived the operator grouping and extended alphabet methods, developed the RAMPS procedure, and wrote the initial code base. AAS developed the optimized hash map routines, created the fermionic caching optimization methods, devised the efficient upper / lower triangle evaluation, ran the fermionic system examples, and contributed to the overall performance of the framework.  HK implemented and ran the SKQD workflow experiments, as well as helped derive the extended operator functionality and RAMPS. All authors contributed to the manuscript.

\bmhead{Funding}

The authors declare no relevant funding.



\bibliography{refs}

\end{document}